\documentclass[3p, preprint, 12pt]{elsarticle}

\usepackage{graphicx}
\usepackage{amssymb}

\usepackage{lineno}

\usepackage{amsmath}
\usepackage{mathtools}
\usepackage{bm}
\usepackage{subcaption}
\usepackage{booktabs}
\usepackage{multirow}
\usepackage{xcolor}
\usepackage[ruled,vlined]{algorithm2e}

\begin{document}

\begin{frontmatter}


\title{Quantifying the uneven efficiency benefits of \\ridesharing market integration}



\author[HKUDUAP]{Xiaohan Wang \fnref{cofirst}}
\author[HKUDUAP,HKUIDS]{Zhan Zhao \corref{cor1} \fnref{cofirst}}
\author[PKUSOG]{Hongmou Zhang}
\author[MITCEE]{Xiaotong Guo}
\author[MITDUSP]{Jinhua Zhao}

\address[HKUDUAP]{Department of Urban Planning and Design, The University of Hong Kong, Hong Kong, China}
\address[HKUIDS]{Musketeers Foundation Institute of Data Science, The University of Hong Kong, Hong Kong, China}
\address[PKUSOG]{School of Government, Peking University, Beijing, China}
\address[MITCEE]{Department of Civil and Environmental Engineering, Massachusetts Institute of Technology, MA, USA}
\address[MITDUSP]{Department of Urban Studies and Planning, Massachusetts Institute of Technology, MA, USA}

\cortext[cor1]{Corresponding author (zhanzhao@hku.hk)}
\fntext[cofirst]{The authors contribute equally to this paper.}

\begin{abstract}
Ridesharing is recognized as one of the key pathways to sustainable urban mobility. With the emergence of Transportation Network Companies (TNCs) such as Uber and Lyft, the ridesharing market has become increasingly fragmented in many cities around the world, leading to efficiency loss and increased traffic congestion. While an integrated ridesharing market (allowing sharing across TNCs) can improve the overall efficiency, how such benefits may vary across TNCs based on actual market characteristics is still not well understood. In this study, we extend a shareability network framework to quantify and explain the efficiency benefits of ridesharing market integration using available TNC trip records. Through a case study in Manhattan, New York City, the proposed framework is applied to analyze a real-world ridesharing market with 3 TNCs---Uber, Lyft, and Via. It is estimated that a perfectly integrated market in Manhattan would improve ridesharing efficiency by 13.3\%, or 5\% of daily TNC vehicle hours traveled. Further analysis reveals that (1) the efficiency improvement is negatively correlated with the overall demand density and inter-TNC spatiotemporal unevenness (measured by network modularity), (2) market integration would generate a larger efficiency improvement in a competitive market, and (3) the TNC with a higher intra-TNC demand concentration (measured by clustering coefficient) would benefit less from market integration. As the uneven benefits may deter TNCs from collaboration, we also illustrate how to quantify each TNC's marginal contribution based on the Shapley value, which can be used to ensure equitable profit allocation. These results can help market regulators and business alliances to evaluate and monitor market efficiency and dynamically adjust their strategies, incentives, and profit allocation schemes to promote market integration and collaboration.

\end{abstract}


\end{frontmatter}


\section{Introduction}

The increasing popularity of app-based ride-hailing services has facilitated the growth of many Transportation Network Companies (TNCs), such as Uber and Lyft, over the past decade. As a result, these TNCs often need to compete for market shares in different cities and regions around the world. Healthy market competition is generally desirable, as it incentivizes better service quality and lower prices. However, this can also lead to market fragmentation, especially considering each TNC has its own mobile application or platform for passengers and drivers to interact. For transportation services that heavily depend on economies of scale, market fragmentation may result in significant efficiency losses. This is particularly evident in ridesharing services. Ridesharing allows two or more passengers to dynamically share parts of their trips in a single vehicle, thus reducing vehicle mileage traveled, relieving traffic congestion, and mitigating vehicle emissions \citep{ma_t-share_2013-1}. Studies have concluded that the potential benefits of ridesharing would only materialize in a dense market where a sufficient number of riders are willing to share rides \citep{rodier_dynamic_2016,alonso-gonzalez_what_2021}. However, market fragmentation essentially decreases the density of the demand pool for each TNC. For example, a trip booked on Uber's platform can only be shared with other Uber trips, even though a trip from Lyft may be a better match with a larger vehicle distance reduction. The fragmentation of potentially shareable trips can lead to longer pick-up distances and larger detours due to suboptimal trip matching, limiting the efficiency of ridesharing services.

To quantify the inefficiency incurred from the market fragmentation, previous studies mainly rely on either theoretical market equilibrium models to estimate consumer/operator surpluses and social welfare in ride-hailing services\citep{sejourne_price_2018,frechette_frictions_2019, ZHOU2022103851, zhou_competitive_2022b}, or data-driven market simulations through subsampling taxi trip records \citep{guo_dissolving_2022,vazifeh_addressing_2018,kondor_cost_2022,zhang_economies_2022}. However, only a few studies consider the inefficiency of ridesharing in the fragmented market \citep{guo_dissolving_2022,zhang_economies_2022}, and none of which are based on real-world TNC market fragmentation. In reality, each TNC holds a unique market share with a specific spatiotemporal demand concentration, and thus the potential benefits of ridesharing market integration may vary greatly across space, time, and TNCs. While the effect of spatial unevenness has been discussed through theoretical market simulations with perfect division between two TNCs \citep{zhang_economies_2022}, the same approach cannot be applied to describe the complex spatiotemporal unevenness in the real world with multiple TNCs. Therefore, a more general approach to quantifying the efficiency benefits of ridesharing market integration and its variation across TNCs is needed.

Although market integration would potentially realize substantial efficiency improvement, a monopolistic market can also lead to undesirable consequences in the long run, such as the abuse of market power. To dissolve the inefficiency caused by market fragmentation while ensuring long-term competition, authorities may consider facilitating collaboration among TNCs. Similar to the widely accepted practice of horizontal collaboration in logistics \citep{chabot_service_2018,dai_profit_2012}, a cooperative ridesharing market could allow the alliance to aggregate each TNC's demand and supply information and dispatch all the fleets to serve the demand with a minimum total vehicle hours traveled (VHT). However, the benefits of such collaboration may be unevenly distributed across TNCs, which can pose as a serious barrier to market integration. Therefore, in addition to the TNC-specific benefits, it is also necessary to quantify the marginal contribution of each TNC, so that we can bridge the gap and ensure fairness in the collaborative market.

To address these issues, this study extends a shareability network framework \citep{santi_quantifying_2014} to evaluate ridesharing markets with multiple TNCs. By comparing the travel duration savings in the fragmented and integrated markets, we can estimate the potential efficiency benefits from market integration for the whole market as well as for individual TNCs. In addition, market simulations are conducted to explore how the efficiency benefits vary based on the demand density, market shares, inter-TNC unevenness (i.e., to what extent the spatiotemporal demand concentration varies across TNCs), and intra-TNC unevenness (i.e., to what extent the demand is concentrated in space and time for a specific TNC). Through a case study in Manhattan, New York City (NYC), real-world TNC trip data is used to uncover the actual market fragmentation patterns across three main TNCs---Uber, Lyft, and Via. The empirical results in Manhattan indeed suggest that the potential efficiency benefits from market integration are highly uneven across TNCs. To facilitate collaboration between TNCs, this study further demonstrates how to use the Shapley value to quantify each TNC's marginal contribution, which can serve as the basis for ensuring fair profit allocation in a collaborative market. The developed methodology enables authorities to evaluate and monitor the market efficiency based on the available trip records so that they can dynamically adjust the policies and strategies to improve market efficiency and encourage collaboration.

\section{Literature Review}

\subsection{Market fragmentation and collaboration} \label{lit:methods}
With the rapid growth of TNCs globally over the last decade, the market is gradually fragmented. By 2019, most local markets worldwide were fragmented and contained two or more TNCs, such as Grab and Go-Jek in Southeastern Asia, Uber and Bolt in Europe, and Uber and Lyft in North America \citep{wang_ridesourcing_2019}. Multiple TNCs offering homogenous ridesharing services in the same city would reduce the density of both the demand (i.e., passengers) and supply (i.e., drivers), increasing vehicle mileage and lowering efficiency. 

In transportation services, where economies of scale are often critical, the inefficiency caused by market fragmentation is a common issue, and many strategies have been proposed to address it. For example, in the freight transportation sector, horizontal collaboration has become widespread, allowing different carriers to share information, orders, and capacities. As the central broker, freight forwarding entities seek the equilibrium between the demanded and available capacities across multiple carriers by interchanging customer requests \citep{krajewska_collaborating_2006}. Collaboration is effective in lowering delivery costs, eliminating empty backhauls, raising vehicle utilization rates, enhancing the quality of service, reducing carbon emissions, and realizing a higher profit level \citep{chabot_service_2018,cruijssen_joint_2007,dai_profit_2012}. 

Collaboration in ride-hailing services is emerging. Some mobile map applications have started to play the integrator role by aggregating the information across multiple TNCs. It allows users to make requests to more than one TNCs simultaneously through the mobile application and match with available drivers affiliated with any TNCs \citep{zhou_competitive_2022b}. Scholars have also proposed to introduce a third-party central broker into the cooperative market \citep{guo_dissolving_2022,kondor_cost_2022,zhou_competitive_2022b}. The third-party entity could be governmental authorities, non-governmental organizations, or private entities. \cite{kondor_cost_2022} claimed that passenger demand information should be aggregated to a collector through an open API of a third-party mobile application. \cite{guo_dissolving_2022} proposed a theoretical framework of four cooperative market designs. One of them is the shared mobility marketplace, where the TNCs need to purchase the requested information from the central broker. 

However, the difficulties of collaboration lie in two aspects. First, TNCs may not readily cede access to passenger information to third parties. Second, TNCs do not benefit evenly from the collaboration. By simulating an uneven market split among three TNCs, \cite{kondor_cost_2022} pointed out that market integration would save more travel costs for operators with small market shares than those with substantial market shares. While smaller TNCs would reach more potential passengers from the integrated platform, those TNCs that already have a large number of users would have few to gain, and thus may lack incentives to participate in the collaboration. Supposedly, their contributions should be considered when sharing the total profits \citep{frisk_cost_2010,karam_horizontal_2021}. The profit-sharing scheme in the cooperative alliance is essential to maintain TNCs' willingness to cooperate and the alliance's stability, but this has rarely been discussed.

\subsection{Inefficiency in the fragmented market}

Previous works on the inefficiency of fragmented transportation markets are summarized in Table~\ref{table:literature}. Existing methods adopted in these studies can be generally categorized into two types. On the one hand, market equilibrium models are often used to simulate the interaction between supply and demand of ride-hailing services, based on which we may evaluate the inefficiency from the aspects of platform profits, consumer surplus, and social welfare. \cite{sejourne_price_2018} built a demand model to confirm that demand fragmentation across multiple platforms could indeed result in higher operating costs and VMT. \cite{frechette_frictions_2019} developed a dynamic equilibrium model to analyze the matching friction, and found that market fragmentation could lead to a lower market thickness and longer waiting and cruising time, generating lower total consumer surplus and lower driver revenue compared to the monopolistic market. \cite{ZHOU2022103851, zhou_competitive_2022b} compared the system performances in the fragmented and integrated platforms at the Nash equilibrium and social optimum. It showed that the integrated platform contributes to higher total realized demand and lower trip fares. They further quantified the inefficiency of the fragmented market in terms of the ratio of social welfare at social optimum to that at Nash equilibrium and found that it decreases as the number of platforms increases.

On the other hand, some studies adopt a graph-based approach to encoding the spatiotemporal relationship between trips. Using Manhattan's yellow taxi trip data, \cite{santi_quantifying_2014} first introduced a shareability network to assess the ridesharing efficiency. Several later works have adapted this framework and datasets to analyze the inefficiency by comparing the travel costs between the fragmented and integrated markets \citep{guo_dissolving_2022,zhang_economies_2022}. \cite{guo_dissolving_2022} found that the duopoly market with a 50/50 market split would have 5\% more vehicle miles traveled (VMT) than an integrated market. In addition, \cite{vazifeh_addressing_2018} applied a vehicle-sharing network to estimate the minimum fleet size needed to serve all taxi trips. A smaller fleet size indicates higher vehicle utilization, less cruising time, and more efficient operations. Their study revealed that the increases in operators could lead to a 4-6\% increase in the total fleet size for a duopoly market and about a 6-10\% increase for a tripoly market. \cite{kondor_cost_2022} further developed this method and applied it to five cities to examine the cost of non-coordination among TNCs. Based on simulations using taxi trip data, they found that each additional TNC entering the market requires an increase in the fleet size ranging from 2.5\% to 67\%. 

\begin{table}[ht!]
  \centering \footnotesize
  \caption{Summary of previous studies on inefficiency incurred in fragmented markets}
  \resizebox{\linewidth}{!}{%
    \begin{tabular}{p{1.7cm}p{1.7cm}p{2cm}p{2.5cm}p{5cm}p{3cm}}
    \hline
    Study & Ridesharing & Method & Datasets & Results &  Critical factors\\
    \hline
    \cite{guo_dissolving_2022} & \checkmark & Shareability network & Yellow taxi trip data in NYC 2011 & A 50/50 market would have 5\% more VMT than an integrated market in ridesharing services.	&  Trip density, sharing thresholds\\
        \hline
    \cite{zhang_economies_2022} & \checkmark & Shareability network & Yellow taxi trip data in NYC 2011 & Four factors associated with efficiency loss in the fragmented ridesharing services are identified.& Trip density, sharing thresholds, market share, spatial unevenness  \\
        \hline
    \cite{vazifeh_addressing_2018} & -- & Vehicle-sharing network & Yellow taxi trip data in NYC & The duopoly market would increase the fleet size by about 4-6\%, while the tripoly market increase by about 6-10\%. & -- \\
        \hline
    \cite{kondor_cost_2022} & -- & Vehicle-sharing network & Taxi trip data in five cities for different years & Due to non-coordination among operators, each additional TNC can increase up to 67\% of fleet size. &  Market share, city-specific traffic speed, trip density \\
        \hline
    \cite{sejourne_price_2018} & -- &Equilibrium model & Yellow taxi trip data in NYC 2017 & Competition among multiple operators would benefit consumers but lead to a significant operational cost loss.& -- \\
        \hline
    \cite{frechette_frictions_2019} & -- & Equilibrium model& Yellow taxi trip data in NYC & Market segmentation reduces market thickness and increases searching time, generating lower total consumer surplus, lower driver revenue, and more waiting time. &  Density of spatial market \\
        \hline
    \cite{ZHOU2022103851} & -- & Equilibrium model & -- & The inefficiency ratio decreases as the number of platforms increases. & Number of platforms \\
    \hline
    \cite{zhou_competitive_2022b} & -- & Equilibrium model & -- & Integration can increase total realized demand and social welfare at both Nash equilibrium and social optimum. & Number of platforms, fleet size  \\
    \hline
    \end{tabular}}
  \label{table:literature}%
\end{table}%

Several factors have been found to be related to the inefficiency in the fragmented market. First, trip density has been recognized as a critical factor in previous studies. It is defined as the number of trips per day per square kilometer (km$^2$) \citep{kondor_cost_2022}, or a relative proportion based on the total number of taxi trips \citep{santi_quantifying_2014,frechette_frictions_2019,zhang_economies_2022}. \cite{santi_quantifying_2014} found that the near-maximum shareability could be reached with 25\% of the trip in Manhattan. Using the same data, \cite{zhang_economies_2022} showed that the total travel cost can be approximated with a simple power function of the trip density. Second, some studies also explored the impacts of market fragmentation level in terms of market share splits between two TNCs \citep{kondor_cost_2022,zhang_economies_2022} or the number of platforms \citep{ZHOU2022103851,zhou_competitive_2022b}. There is little discussion on how the various market split among multiple TNCs affects ridesharing efficiency. Third, the setting of sharing thresholds (e.g., the largest acceptable delay for ridesharing passengers) can also influence the efficiency \citep{santi_quantifying_2014,zhang_economies_2022}. In addition, \cite{kondor_cost_2022} suggested that the travel speed in the city can affect the cost of non-coordination between TNCs through a city-specific constant. \cite{zhou_competitive_2022b} also demonstrated the impact of the fleet size and commission fee of the integrators in the analysis of profit and welfare. 
However, few studies have considered the spatiotemporal unevenness of demand concentration and its variation across TNCs. \cite{zhang_economies_2022} identified four influencing factors for the efficiency of a fragmented market and, for the first time, discussed the potential effect of spatial concentration; more significant spatial overlapping of demand between two TNCs would lead to larger inefficiency. However, both the spatial and temporal distributions of TNC trips should be considered to characterize uneven demand concentration in the real world, a more generalizable measurement of unevenness across multiple TNCs is needed.

\subsection{Research gaps} 
\label{sec:GCN_review}

To summarize, the research gaps mainly lie in three aspects. First, very limited works have attempted to quantify the potential efficiency benefits generated from market integration for ridesharing services, and none of them are based on a real TNC market fragmentation. The simulated results based on the taxi trip data may not reflect the actual market fragmentation among TNCs across time and space, as the heterogeneous demand concentration patterns of TNCs are ignored. This leads to the second gap that the impact of the spatiotemporal unevenness in demand concentration has not been discussed. The unevenness can come from the demand concentration within a specific TNC (intra-TNC unevenness) and the concentration variances across different TNCs (inter-TNC unevenness). Third, previous simulated results have shown that market integration would benefit smaller TNCs more significantly, leading to unfairness and hindering market collaboration. To encourage market integration through collaboration, we need to better understand how the benefits are unevenly distributed among TNCs and how to design profit-sharing schemes to ensure fairness. This study aims to address all three research gaps by analyzing a real-world fragmented ridesharing market based on publicly available TNC trip records, considering spatiotemporal unevenness through shareability networks, and applying the shapely value to quantify TNC-specific contributions.

\section{Methods} 

\subsection{Study framework}

\begin{figure}[!ht]
  \centering
  \includegraphics[width=0.8\textwidth]{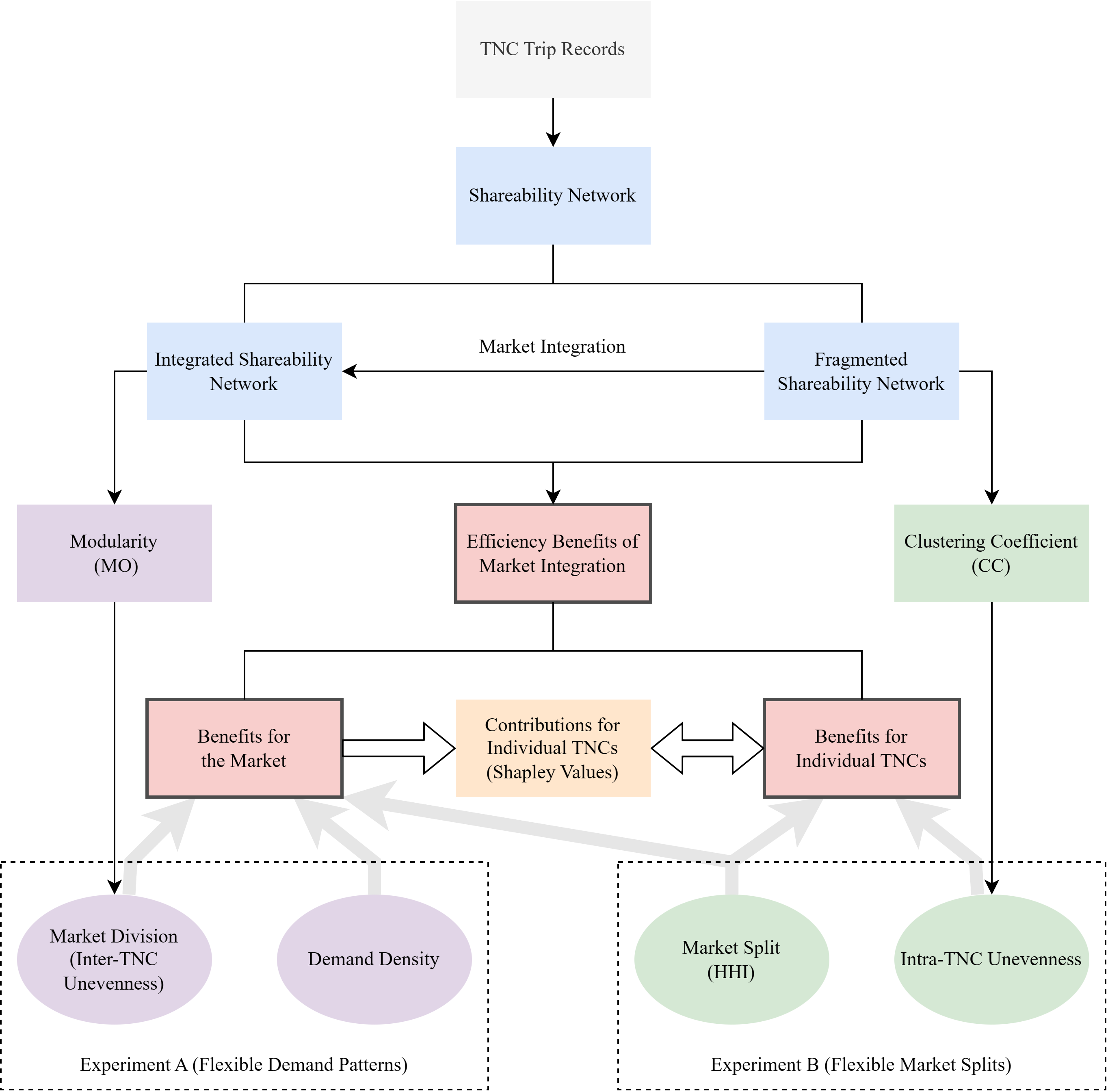}
  \caption{Study framework}\label{fig:frame}
\end{figure}

The overall research framework of this study is summarized in Fig.~\ref{fig:frame}. To estimate ridesharing efficiency, we first build a shareability network based on TNC trips, where each node is a trip, and each edge indicates the shareability between two trips \citep{santi_quantifying_2014}. If two trips can be served by the same vehicle without incurring too much detour, they are regarded as shareable. The efficiency of sharing two trips can be quantified as the vehicle travel duration saved by serving the two trips with one vehicle instead of two separate vehicles. In this study, we use the vehicle hours traveled (VHT) saving to measure the ridesharing efficiency. Given a shareability network, we can estimate its potential ridesharing efficiency by first identifying the optimal matching between shareable trips and then summing up the VHT savings of the matched trips. In a fragmented market, only two trips associated with the same TNC can be shared, and thus each TNC has its own shareability network. In an integrated market, however, any two trips can be shared, leading to a single combined shareability network. In this study, the \textit{efficiency benefit} of ridesharing market integration is defined as the percentage increase of the VHT saving in the integrated market compared to the fragmented market. Since the overall efficiency benefit may not be evenly distributed across TNCs, their individual benefits should be further evaluated.

To understand how the efficiency benefit of ridesharing market integration varies based on the market conditions, we further conduct two experiments to simulate different market scenarios. In Experiment A (\textit{Flexible Demand Patterns}), we simulate different levels of demand density and concentration based on the current market shares. We first discuss how the efficiency benefit may change when the demand density decreases due to lower passenger willingness to share in Experiment A1. Moreover, the spatiotemporal distributions of demand are uneven across TNCs due to their distinct business strategies. If the spatiotemporal demand concentration varies significantlly across TNCs, it suggest a stronger market division. In this study, we use the modularity of the combined shareability network to measure the inter-TNC unevenness of demand concentration in space and time, which reflects the degree of market division. To investigate the effect of market division, we simulate different levels of inter-TNC unevenness in Experiments A2 and A3 and examine how the efficiency benefit varies.

Experiment B (\textit{Flexible Market Splits}) focuses on scenarios when the market competition intensity varies. In this experiment, we keep the demand density constant and simulate markets with different market splits. Specifically, Herfindahl-Hirschman Index ($HHI$) is used to describe the market split and incurred competition level, and its relationship with the efficiency benefit of market integration is investigated. While the overall efficiency benefit is quantified for the whole market, individual TNCs are more concerned about their individual benefits. Therefore, we further discuss the influencing factors for the efficiency benefits of market integration for each TNC, including the effect of intra-TNC unevenness, measured as the clustering coefficient of TNC-specific shareability network. 

To facilitate collaboration, each TNC should receive a fair share of the total benefit in market integration that matches its contribution. To address this issue, this study also demonstrates the use of the Shapley value to measure the expected marginal contributions of individual TNCs, which can serve as the basis for allocating the profit generated through collaboration.

\subsection{Shareability network}

In this study, we extend the shareability network approach \citep{santi_quantifying_2014} to quantify the potential ridesharing efficiency. The shareability network transforms the spatiotemporal distribution of trips into a theoretic graph $G = (V,E)$, where $V$ is the set of trips and $E$ the set of edges that indicates the shareability between trips. Following the original work, we assume that at most two trips can be shared. Two trips are shareable if serving them together with one vehicle can (1) generate total travel duration savings and (2) only incur an acceptable delay for the involved passengers.

Each trip $x\in V $ can be represented as a vector $(o_x,d_x,t_x^o,t_x^d)$, where $o_x$ denotes the origin of trip, $d_x$ the destination, $t_x^o$ the trip start time, and $t_x^d$ the trip end time. $c(x)$ denotes the travel time of the trip, or $c(x)=t_x^d- t_x^o$. To identify shareable trips, we use a simple rerouting process. Each pair of trips $i$ and $j$ can be combined by connecting the two origins and destinations in four possible orders, including $o_i\rightarrow o_j\rightarrow d_i\rightarrow d_j$, $o_i\rightarrow o_j\rightarrow d_j\rightarrow d_i$, $o_j\rightarrow o_i\rightarrow d_i\rightarrow d_j$, and $o_j\rightarrow o_i\rightarrow d_j\rightarrow d_i$, provided that the two trips overlap in start and end times. The ridesharing service is viable for both the operator and whole society only when it generates travel duration savings compared to the non-shared services. Hence, to satisfy condition (1), the duration of the combined trip $T_{ij}$ cannot exceed the sum of two single trips, or $c(T_{ij} )< c(i)+c(j)$.

For condition (2), we define a maximum delivery delay $\mu$ to indicate the maximum acceptable travel delay for ridesharing passengers, including the pickup and in-vehicle delays. For the combined trip $T_{ij} $, $T_{ij}(o_x)$ is the pickup time at $o_x$, and $T_{ij}(d_x)$ is the drop-off time at $d_x$ in the combined trip. For $T_{ij} $ to be feasible, it needs to satisfy the following:

\begin{equation}
\begin{cases}
  \ t_i^o\leq T_{ij}(o_i)\leq t_i^o+\mu;\\
  t_j^o\leq T_{ij}(o_j)\leq t_j^o+\mu;\\
  T_{ij}(d_i)\leq t_i^d+\mu;\\
  T_{ij}(d_j)\leq t_j^d+\mu,
\end{cases}   
\end{equation}

We check the combined routes in four orders. If more than one feasible route exists, the shared route will follow the order with the greatest savings in travel duration. For example, the route $ o_i\rightarrow o_j\rightarrow d_i\rightarrow d_j$ can be verified by checking whether a value of $T_{ij}(o_i)$ that satisfies the following conditions exists.

\begin{equation}
\begin{cases}
  \ t_i^o \leq T_{ij}(o_i) \leq t_i^o+\mu;\\
  t_j^o\leq T_{ij}(o_i)+c(T_{o_i o_j}) \leq t_j^o+\mu;\\
  T_{ij}(o_i)+ c(T_{o_i o_j})+ c(T_{o_j d_i})\leq t_i^d+\mu;\\
  T_{ij}(o_i)+ c(T_{o_i o_j})+ c(T_{o_j d_i})+ c(T_{d_i d_j})\leq t_j^d + \mu,
\end{cases}   
\end{equation}

To estimate the travel duration of rerouted trips, we can either directly query online mapping services, or compute it based on the real-world street network that is widely available. In this study, we adopt the latter approach, due to the large number of rerouted trips to analyze. Specifically, the street network is regarded as a directed weighted graph, where each street link is an edge, and each intersection is a node. The travel duration for passing through an edge is recorded as the weight. Based on the street network, the Dijkstra algorithm is applied to compute the shortest path of rerouted trips and calculate the travel time in four different orders. For simplicity, we assume constant speed, which can be calibrated based on observed trip records or external speed records. If detailed traffic state data is available, the network weights can be treated as dynamic, but the general approach should still work.

\begin{figure}[!ht]
  \centering
  \includegraphics[width=0.7\textwidth]{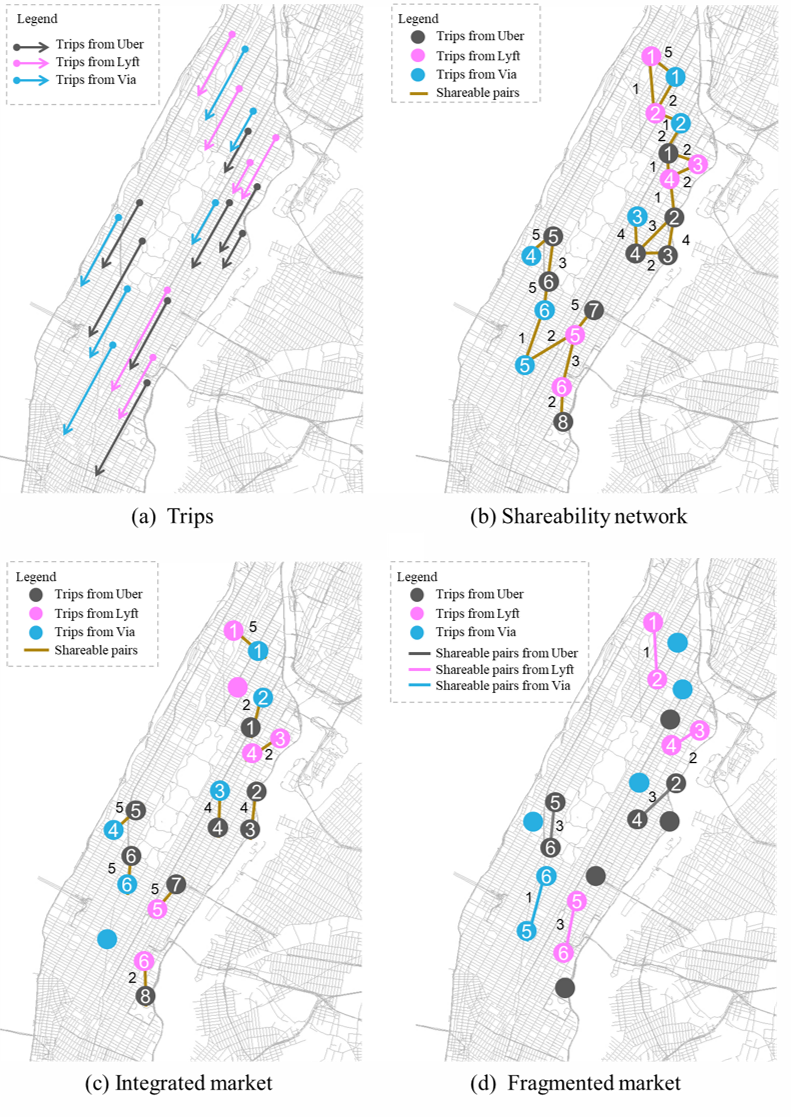}
  \caption{Illustration of the shareability network and maximum weighted matching in both integrated and fragmented markets in New York City with 3 TNCs---Uber, Lyft, and Via (adapted from \cite{santi_quantifying_2014})}\label{fig:ill}
\end{figure}

With the shareable trips identified, an undirected weighted shareability network $G$ can be constructed, where each trip is considered as a node. If two trips are shareable, an edge will connect them. For any pair of trips $(i,j)$, the VHT savings by serving a shared trip $T_{ij}$ compared to the two non-shared trips will be assigned to the edge as its weight, or $w_{ij}=c(i)+c(j)-c(T_{ij})$. A higher weight indicates more VHT saved by sharing the trips $(i,j)$. As shown in Fig.~\ref{fig:ill}(b), the shareability network connects the shareable trips with edges, whose weights (numbers in black color) represent the potential travel time savings by sharing the pair of trips.

\subsection{Efficiency benefits for the market and individual TNCs}

This study aims to estimate the efficiency benefit of ridesharing market integration, measured as the percentage increase of VHT saving, both for the overall market and individual TNCs. The shareability network allows a trip to be shareable with multiple other trips. However, in the actual implementation of ridesharing services, we assume that only two trips can be shared. To determine which two trips should be matched together, we follow \cite{santi_quantifying_2014} to adopt the maximum weighted matching algorithm to find an optimal matching $M$ in $G$, so that the total weights of the edges in $M$ is maximized \citep{galil_efficient_1986}. $M$ is a special graph, which shares the same nodes as $G$ but only has a subset of edges from $G$. Specifically, no node in $M$ can be connected to more than one other node, and thus $M$ represents a trip matching result, such as the examples shown in Fig.~\ref{fig:ill}(c-d). The maximum weighted matching algorithm finds the optimal $M$ that maximizes the cumulative VHT saving, indicating the ridesharing efficiency:

\begin{equation}
\label{eq:VHTsaving}
\max_{M} VHT_{M} = \sum_{(i ,j) \in M} w_{ij}
\end{equation}

The maximum weighted matching algorithm can be applied in integrated and fragmented markets. As illustrated in Fig.~\ref{fig:ill}(c), the integrated market allows the TNCs to cooperate and pool the trips together when searching for matches among passengers. Fig.~\ref{fig:ill}(d) shows that the market fragmentation breaks the edges that connect trips from two different TNCs. In this case, each TNC can only find shareable trips among its own demand pool. As shown in Fig.~\ref{fig:ill}(c-d), the integrated market could generate more shareable trips and VHT savings than the fragmented case. 

Let us use $M_{I}$ to denote the optimal matching in the integrated market. For the fragmented market contains multiple TNCs $N$, each TNC forms a shareability network $G_1,G_2,...,G_{|N|}$. Then the maximum weighted matching algorithm can be used to compute the optimal matchings for different TNCs in the fragmented market, $M_{F}=\{M_1,M_2,...,M_{|N|} \}$. Finally, we calculate the VHT savings both in the fragmented and integrated markets. The efficiency benefit $Q$ of ridesharing market integration is defined as the difference in VHT savings between the integrated and fragmented markets divided by the VHT saving in the fragmented market: 

\begin{equation}
\label{eq:Q}
Q = \frac{\sum_{(i ,j) \in M_{I}} w_{ij} -\sum_{(i ,j) \in M_{F}} w_{ij} }{\sum_{(i ,j) \in M_{F}} w_{ij} } \times 100\%
\end{equation}

From the perspective of individual TNCs, the integration of ridesharing market may result in different levels of efficiency improvement for them. To capture such variation, the individual benefit $Q_p$ for a TNC $p$ in the integrated market is quantified as the increase in the percentage of potential VHT savings in ridesharing, indicating the extent to which market integration can improve their ridesharing efficiency. For TNC $p$ in the fragmented market, $V_p$ denotes the set of shareable trips and $M_p$ is the optimal matching for shareable trips. The VHT savings of TNC $p$ in the fragmented market is the sum of the edge weights in $M_p$. In the integrated market, if a shared trip involves a pair of trips from different TNCs, we assume that each TNC would gain half of the VHT saving incurred by ridesharing regardless of who the carrier is in actual operation. Let $M_{in}^p$ be a subset of $M_{I}$, containing the edges that connect two nodes from $V_p$, and $M_{out}^p$ be a subset containing the edges that connect one node from $V_p$ and the other from a different TNC. The VHT saving of TNC $p$ in the integrated market can be calculated by adding up the cumulative weights of edges in $M_{in}^p$ and half of the cumulative weights of edges in $M_{out}^p$. Finally, the increase in VHT saving is calculated by subtracting the potential VHT saving in the fragmented market from the saving in the integrated market. Therefore, the individual benefit $Q_p$ for TNC $p$ from market integration can be expressed as:

\begin{equation}
\label{eq:Qp}
Q_p = \frac{\sum_{(i ,j) \in M_{in}^p} w_{ij} +0.5*\sum_{(i ,j) \in M_{out}^p} w_{ij} -\sum_{(i ,j) \in M_p} w_{ij} }{\sum_{(i ,j) \in M_p} w_{ij} } \times 100\%
\end{equation}

\subsection{Measures of spatiotemporal unevenness}

Trip density and market shares have been identified in the literature as two main factors influencing the fragmented market efficiency. However, both of them are based on the number of trips and fail to consider the variability in the spatiotemporal distribution of these trips. In this study, two additional measures are used to examine how the spatiotemporal unevenness of trip concentration affects the efficiency benefit for the whole market and individual TNCs.

We propose to use the modularity of the integrated shareability network to measure the inter-TNC unevenness in the spatiotemporal distribution of trips. Modularity is commonly used in community detection, a fundamental task in network science. A larger value of modularity suggests a stronger community structure in the sense that there are many edges within communities and only a few between them \citep{newman_modularity_2006}. Adopting this concept, each TNC is considered as a community in the integrated shareability network. When TNCs have distinct demand concentrations with little overlap in space or time, there would be more shareable trips within each TNC, and fewer between TNCs. In such cases, the integrated market exhibits significant market division with a greater value of modularity.

Given an integrated market formed by multiple TNCs $G_I$, its modularity $MO$ can be calculated as

\begin{equation}
\label{eq:mo}
MO =\sum_{p=1}^{|N|} \bigg[\frac{|E_{in}^p|}{|E_{I}|} -\bigg(\frac{\sum_{v\in V_p}deg(v)}{2|E_{I}|} \bigg)^2 \bigg]
\end{equation}
where $E_{I}$ indicates the set of edges in the integrated shareability network $G_I$, $E_{in}^p$ is the subset of $E_{I}$ that connect two trips belonging to TNC $p$, $V_p$ is the set of trips for TNC $p$, and the degree of a trip $deg(v)$ is the number of edges that are incident to the node, representing the number of shareable trips. A larger value of $MO$ indicates a more pronounced market division among TNCs in time and space. Usually, a modularity value exceeding 0.3 is considered a significant division \citep{clauset_finding_2004}. 
A negative value represents that the demands from different TNCs overlap spatiotemporally and do not show apparent market division. 

We then use the clustering coefficient $(CC)$ to describe the intra-TNCs unevenness in the spatiotemporal distribution of trips. The clustering coefficient is a widely-used measure of the extent to which nodes in a graph tend to cluster together. The local clustering coefficient for a node quantifies how close its neighbors are connecting each other \citep{watts_collective_1998}. It is calculated by dividing the number of edges between a node's neighbors by the number of possible edges between neighbors. In ridesharing, $CC$ measures the extent to which the trips are spatiotemporally close to each other, reflecting the shareability of trips. A given number of trips could have different levels of shareability, which largely depends on the spatiotemporal concentration of trips. For an individual TNC, the shareability would be greater when trips tend to cluster together in space and time, and smaller when the trips are evenly and sparsely distributed. In this study, we measure the network average clustering coefficient of TNC $p$ as

\begin{equation}
\label{eq:cc}
CC_p = \frac{1}{|V_p|} \sum_{v \in V_p} \bigg(\frac{2R(v)}{deg(v)(deg(v)-1)} \bigg)
\end{equation}
where $R(v)$ is the number of triangles through node $v$.

\subsection{Scenario Experiments}

Two experiments are conducted to explore how the efficiency benefit of market integration would vary based on various market characteristics, including demand density, market shares, inter-TNC spatiotemporal unevenness, and intra-TNC spatiotemporal unevenness. In Experiment A (\textit{Flexible Demand Patterns}), we investigate the effect of demand distribution, where we subsample trips to create markets with different levels demand density and inter-TNC unevenness, with a fixed market split. This experiment simulates how the efficiency benefit would change with passengers' willingness to share or the market division between TNCs.

Under Experiment A, we consider three variations. In Experiment A1, we randomly subsample trips to form a series of subgraphs $G_\alpha$. $\alpha$ denotes the demand density levels, ranging from 5\% to 100\% with 5\% intervals. For example, $G_{90\%}$ represents a market containing 90\% of trips from each TNC. In doing so, we examine how the efficiency benefit varies with the demand density. 

To simulate markets with diverse inter-TNC spatiotemporal unevenness, we repeat Experiment A1 by subsampling trips from demand-sparse and demand-dense spatiotemporal regions in Experiments A2 and A3, respectively. For each TNC, we assume that trips with a high shareability network degree are from demand-dense regions, while trips with low degrees are from demand-sparse regions. Specifically, trips are ranked based on their degrees in the fragmented shareability network for each TNC. Experiment A2 chooses some proportion of low-degree trips for each TNC to simulate a series of markets with demand more sparsely and evenly distributed in time and space. Experiment A3 organizes markets by only keeping the high-degree trips in each TNC. It is expected that, while the concentrations of dense demands are more distinct and TNC-specific, the sparsely scattered demands tend to share similar spatiotemporal distributions. Therefore, for the market with specific trip density $\alpha$, the market division in Experiment A3 is likely to be more significant (i.e., more inter-TNC unevenness) than Experiment A1, and the market division in Experiment A2 would be less significant (i.e., less inter-TNC unevenness). By comparing the efficiency benefits in Experiments A2 and A3 against those in Experiment A1 with the same demand density level $\alpha$, we can uncover the impact of market division on ridesharing efficiency.

To investigate the effect of market competition levels, Experiment B (\textit{Flexible Market Splits}) keeps the overall market demand density constant and simulates TNCs seizing different market shares. Individual TNCs may compete for a larger market share or concentrate on targeted operational zones to obtain higher returns. This experiment discusses how the market share split and intra-TNC spatiotemporal concentration of trips may affect the efficiency benefits for the whole market as well as individual TNCs. Previous studies on the impact of market split on ridesharing efficiency are mostly limited to duopoly markets. In this study, we use the Herfindahl-Hirschman Index $(HHI)$ to quantify the market split among multiple TNCs. $HHI$ is the sum of the squared market shares of each competing firm \citep{rhoades_herfindahl-hirschman_1993,rhoades_market_1995}, which can be expressed as:

\begin{equation}
\label{eq:hhi}
HHI = \sum_{p=1}^{|N|} MS_p ^2
\end{equation}
where $N$ is the number of TNCs in the fragmented market, $MS_p$ denotes the market share of each TNC $p$. The index perfectly describes the market split by reflecting both the variance in market share inequality and the number of competitors. $HHI$ decreases in value as the market shares of firms become more even or as the number of firms in the market increases \citep{hannan_market_1997}, which is a sign of the market getting competitive. It has also been used to quantify the market competition intensity of the fragmented ride-hailing services \citep{huang_understanding_2022}.  In Experiment B, we simulate the relative market shares of the TNCs by randomly assigning trips, with various market shares for each TNC. We then investigate the relationship between market efficiency and $HHI$ in all possible market share permutations. Also, based on Eq.~\eqref{eq:Qp}, we calculate the individual efficiency benefit for each TNC in all simulated markets to explore how it relates to the intra-TNC unevenness.

\subsection{Shapely value for profit sharing across TNCs}	

A fair profit allocation scheme should consider the characteristics of the participating parties, including their market shares, individual benefits, as well as their individual contributions to the collaboration synergies. To encourage market integration for higher ridesharing efficiency, this study proposes a novel use of the Shapley value for profit-sharing across TNCs. The Shapley value \citep{shapley_shapley_1988} is a widely accepted profit allocation scheme in collaborative games. It is often used in situations where each participant's contribution is unequal, such as profit allocation among carriers in collaborative logistics \citep{dai_profit_2012,kimms_shapley_2016,shi_method_2020}. The Shapley values represent the expected marginal contributions of participants \citep{shapley_shapley_1988}. A fair collaboration requires the profits to be allocated to the participants based on their marginal contribution.

Given a collaboration $(N,val)$, where $N$ is the set of TNCs in the collaboration, $S$ is the alliance formed by any set of TNCs, and $val(S)$ denotes the total expected value that an alliance $S$ can obtain through cooperation, measured by additional VHT savings through the alliance. The Shapley value of TNC $p$ is given as:

\begin{equation}
\label{eq:Shapley}
\phi_p (N,val)=\frac{1}{|N|!}\sum_{S\subseteq N/\{p\}}\left(|S|!(|N|-1-|S|)![val(S\cup \{p\})-val(S)]\right)
\end{equation}

Based on Eq.~\eqref{eq:Shapley}, we can estimate the expected values for all TNC alliance permutations and calculate the Shapley value for each TNC.

\section{Results} 
\subsection{Data}	
In this study, to capture the market characteristics of real-world market fragmentation, we focus on the ridesharing market in Manhattan, NYC. The Manhattan ridesharing market includes 3 high-volume TNCs---Uber, Lyft, and Via, whose trip records are complied and made publicly available by NYC Taxi \& Limousine Commission\footnote{https://www.nyc.gov/site/tlc/about/tlc-trip-record-data.page}. Each trip record contains the following fields: the license ID of the TNC, the timestamp of the pickup and drop-off, the taxi zone ID of pickup and drop-off locations, etc. The temporal information of pickups and drop-offs is accurate to the second. However, for privacy reasons, the exact geographic coordinates of trip origins and destinations are not provided; instead, each pickup or drop-off location is aggregated to one of the 67 taxi zones. 

To obtain the geographic coordinates of each trip endpoint, we build a street network of Manhattan by extracting the edges and nodes from the open map data provided by NYC Department of City Planning\footnote{https://www.nyc.gov/site/planning/data-maps/open-data/dwn-lion.page}. We randomly assign each trip end location to an intersection of the street network within the corresponding taxi zone. Initially, the speed of the street network is assumed to be the average speed of the TNC trips, calculated by the trip distance divided by duration (3.89 m/s). Based on the street network, we use the shortest path function to estimate the travel duration from the assigned trip origins to destinations. We find that the median value of the estimated travel duration is 14.7 min, slightly shorter than the trip records (15.6 min). Based on this difference, we can further calibrate the speed. By setting the street network speed to 3.46 m/s, we keep the estimation error of median trip duration within 0.02 min.

To analyze the efficiency of the ridesharing market, we will focus on a typical Wednesday on May 15, 2019. After discarding the trips in less than 2 minutes, there are 211,535 TNC trips in total, with key market statistics summarized in Table~\ref{table:characteristics}. Based on the selected dataset, Uber seizes 68.1\% of the market, Lyft and Via account for 18.3\% and 13.6\%, respectively. As a result, $HHI=0.5159$.

\begin{figure}[!ht]
  \centering
  \includegraphics[width=0.85\textwidth]{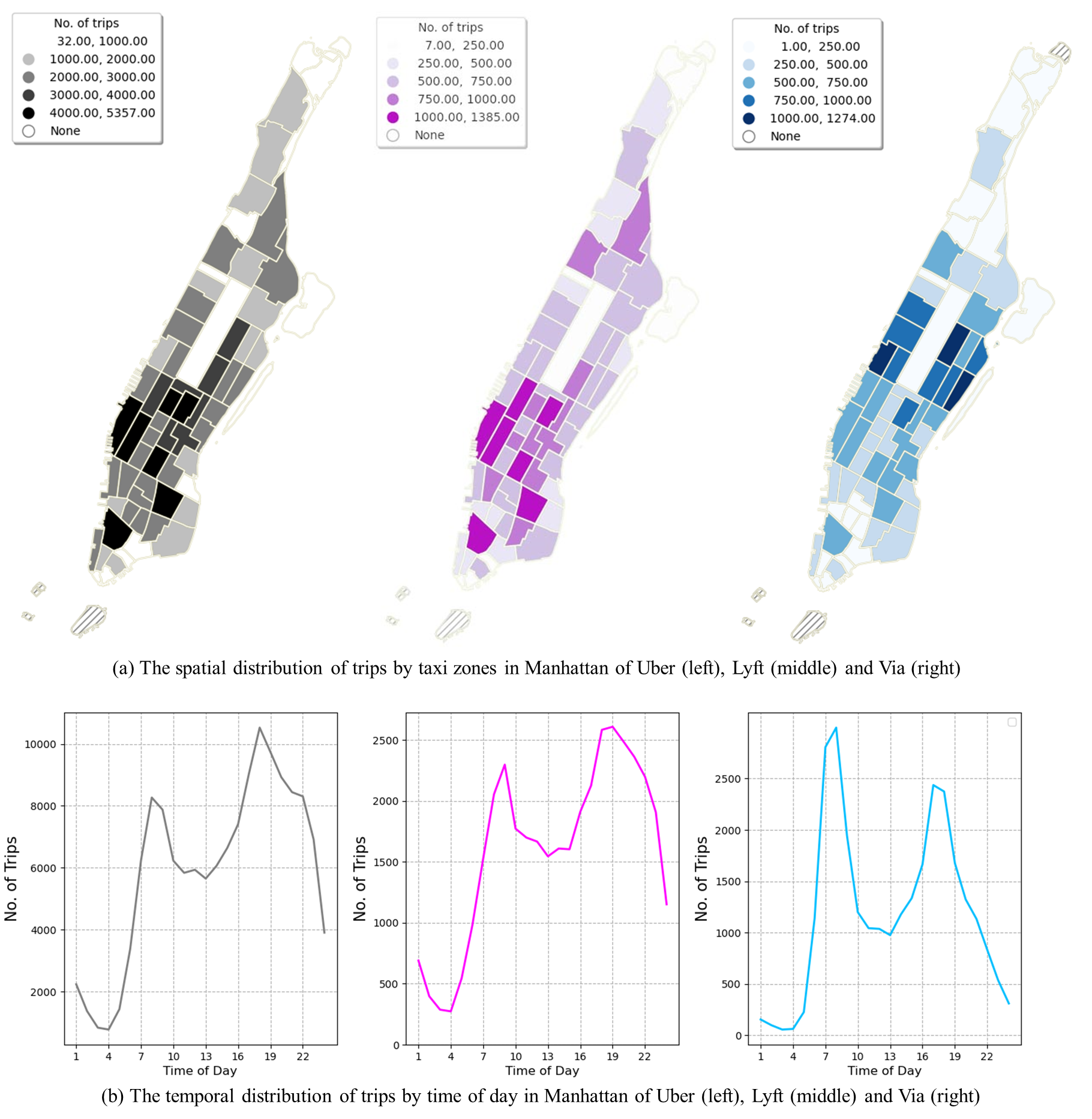}
  \caption{Trip distribution in time and space by TNCs in Manhattan}\label{fig:spatiotemporal}
\end{figure}

In addition, the real-world TNC trip records capture the inter- and intra-TNC unevenness in their spatiotemporal characteristics, resulting from diverse operational, pricing, or marketing strategies by different TNCs. On the one hand, we can identify some variation in spatiotemporal demand patterns across TNCs. Fig.~\ref{fig:spatiotemporal}(a) shows the spatial distribution of trip pickup locations of different TNCs. Uber trips are more evenly distributed across taxi zones. Lyft displays a similar concentration pattern with Uber in the Midtown and Downtown areas. Via focuses more on Uptown areas and has less overlap with the concentration areas of Uber and Lyft. As for the temporal distribution in Fig.~\ref{fig:spatiotemporal}(b), among the three TNCs, Via's peak hours match the typical commuting hours, suggesting a greater portion of commuting trips in Via's demand pool. For Uber and Lyft, in addition to commuting trips, their busiest hours occur in the evening. It implies that Uber and Lyft serve more recreational trips at night than Via. On the other hand, the TNCs show different extents of intra-TNC demand clustering. Uber is supposed to have denser clustering for its much larger market share across most areas and time periods. Via has the sharpest curve in temporal distribution, with most of its demand clustering during commuting hours.

\begin{table}[ht!]
  \centering \footnotesize
  \caption{Market characteristics in Manhattan}
    \begin{tabular}{cccc}
    \addlinespace
    \hline
    Characteristics & Uber & Lyft & Via \\
    \hline
    Demand Density & 144077 & 28663 & 38795\\
    \cmidrule(lr){2-4}
    Modularity & & 0.0098  \\
    \cmidrule(lr){2-4}
    Market Share & 68.1\% & 18.3\% & 13.6\%\\
    \cmidrule(lr){2-4}
    Herfindahl-Hirschman Index $(HHI)$ & & 0.5159 \\
    \cmidrule(lr){2-4}
    Clustering Coefficient & 0.2180 & 0.1894 & 0.2050\\
    \hline
\end{tabular}
\label{table:characteristics}%
\end{table}%

\subsection{Overall efficiency benefits of market integration}

Based on the trip records and street network in Manhattan, we build the shareability networks for the integrated and fragmented markets. To identify shareable trips, a large maximum delivery delay will increase the ridesharing opportunities but compromise the quality of services by generating more passenger waiting time. This study assumes the maximum delivery delay $\mu$ to be 300 seconds, which has been used in previous studies \citep{santi_quantifying_2014,tafreshian_trip-based_2020,zhang_economies_2022}. With the integrated shareability network, we can use modularity to measure the extent of inter-TNC spatiotemporal unevenness, with a value of 0.0098. It indicates that the market division among the 3 TNCs in Manhattan is relatively weak. Similarly, based on the fragmented shareability networks, we can use the clustering coefficient to measure the intra-TNC unevenness, with Uber at 0.2180, Lyft at 0.1894, and Via at 0.2050. This suggests that Uber has the most pronounced spatiotemporal demand concentration, with Via coming in second and Lyft having the sparsest demand.

With the current market fragmentation and demand concentration levels, we find that market integration would potentially generate a 13.3\% efficiency benefit (measured as the percentage of additional VHT savings) in Manhattan. Specifically, 2,607 VHT can be saved, accounting for 5\% of Manhattan's total TNC daily vehicle travel time. Therefore, integrating ride-hailing services shows great potential in easing congestion and reducing vehicle emissions.

\begin{figure}[!ht]
  \centering
  \includegraphics[width=0.95\textwidth]{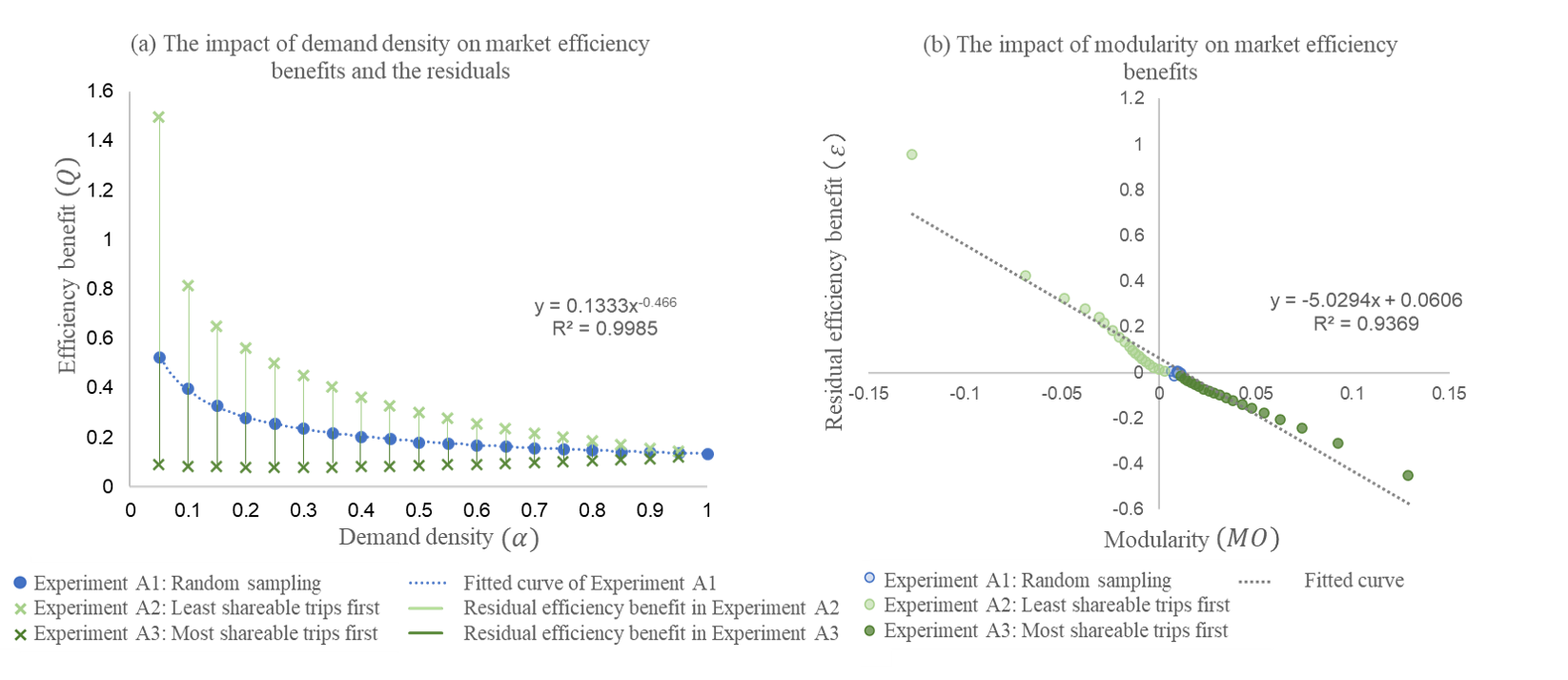}
  \caption{The impact of demand density and inter-TNC unevenness on the market efficiency benefits in integration}\label{fig:demand_division}
\end{figure}

Note that the above efficiency benefit is estimated based on the assumption that all passengers are willing to share their rides. In reality, the willingness to share varies. Through trip subsampling, Experiment A simulates alternative markets with different levels of willingness to share, in which a larger $\alpha$ indicating a higher willingness to share. As shown in Fig.~\ref{fig:demand_division}(a), the efficiency benefit $Q$ exhibits different trends when the trips are sampled differently. When trips are randomly sampled (Experiment A1), the market integration would generate larger relative benefits as passengers' willingness to share rides decreases. This is consistent with prior findings about the effect of trip density \citep{santi_quantifying_2014,frechette_frictions_2019,kondor_cost_2022,zhang_economies_2022}. When the demand gets denser, the effect of market integration would gradually reduce. 

However, the overall demand density alone cannot fully account for network shareability under the integrated market. The diverging results in Experiments A2 and A3, shown in Fig.~\ref{fig:demand_division}(a), suggest that the inter-TNC unevenness in spatiotemporal demand concentration can also play a role. Therefore, we also consider the situations when the three TNCs adjust their ridesharing strategies on targeted space and time, leading to different levels of market divisions. In Experiment A2, instead of random sampling as in Experiment A1, we sample the least shareable trips (nodes with the lowest degrees in the shareability network) for each TNC. The sampled trips tend to be evenly and sparsely distributed for all TNCs, and thus show no apparent market division. Similarly, in Experiment A3, we sample the most shareable trips (nodes with the highest degrees in the shareability network) for each TNC. Due to the distinct demand concentration for each TNC, the sampled trips tend to show a more significant market division and a higher modularity value. Notably, the dark green ``x'' marks in Fig.~\ref{fig:demand_division}(a) illustrate the results for Experiment A3, and they actually show an upward trend, indicating the efficiency benefit can increase as the trip density increases.

To further investigate the diverging trends of the efficiency benefit relative to trip density, we fit a power function (i.e., $y=ax^b$) describing the impact of demand density on efficiency benefit under Experiment A1, shown as the blue dotted line in Fig.~\ref{fig:demand_division}(a). The function fits the blue dots well, as the $R^2$ reaches over 0.99, though it cannot explain the trends for Experiments A2 and A3. $\hat{Q}_\alpha$ is used to denote the predicted efficiency benefit for each density level $\alpha$ under Experiment A1. Based on the fitted function, we capture the residuals in efficiency benefits by:

\begin{equation}
\label{eq:deltaQ}
\begin{cases}
    \varepsilon_\alpha^{(1)} = Q_\alpha^{(1)}-\hat{Q}_\alpha\\
    \varepsilon_\alpha^{(2)} = Q_\alpha^{(2)}-\hat{Q}_\alpha\\
    \varepsilon_\alpha^{(3)} = Q_\alpha^{(3)}-\hat{Q}_\alpha\\
\end{cases}   
\end{equation}
where $Q_\alpha$ and $\varepsilon_\alpha$ denote the actual efficiency benefit and its residual for a given $\alpha$, and their superscripts $(1)$, $(2)$, and $(3)$ represent Experiments A1, A2, and A3, respectively.
Fig.~\ref{fig:demand_division}(b) shows that the residual efficiency benefit can be linearly explained by the modularity indicating inter-TNC unevenness, with the $R^2$ over 0.93. The residual efficiency benefit decreases as the modularity value increases, suggesting that market integration would lead to less efficiency improvement when the market is distinctly divided. Hence, when the market split held constant, the efficiency benefit of market integration is jointly determined by both the demand density (specified by $\alpha$) and the market division (specified by $MO$). Based on the simulation results for Experiment A in Manhattan, the relationship can be specified as follows:

\begin{equation}
\label{eq:market}
Q=0.1333\alpha^{-0.466}-5.0294MO+0.0606
\end{equation}

This equation can be used to estimate the potential efficiency benefit from integrating markets with different demand concentration patterns, taking into account its inter-TNC spatiotemporal unevenness. Generally, integrating the ridesharing market would not generate appreciable efficiency benefits when passengers have a strong willingness to share rides, or when the spatiotemporal demand concentrations are clearly divided across TNCs.

\subsection{Uneven efficiency benefits for individual TNCs}

\begin{figure}[!ht]
  \centering
  \includegraphics[width=0.95\textwidth]{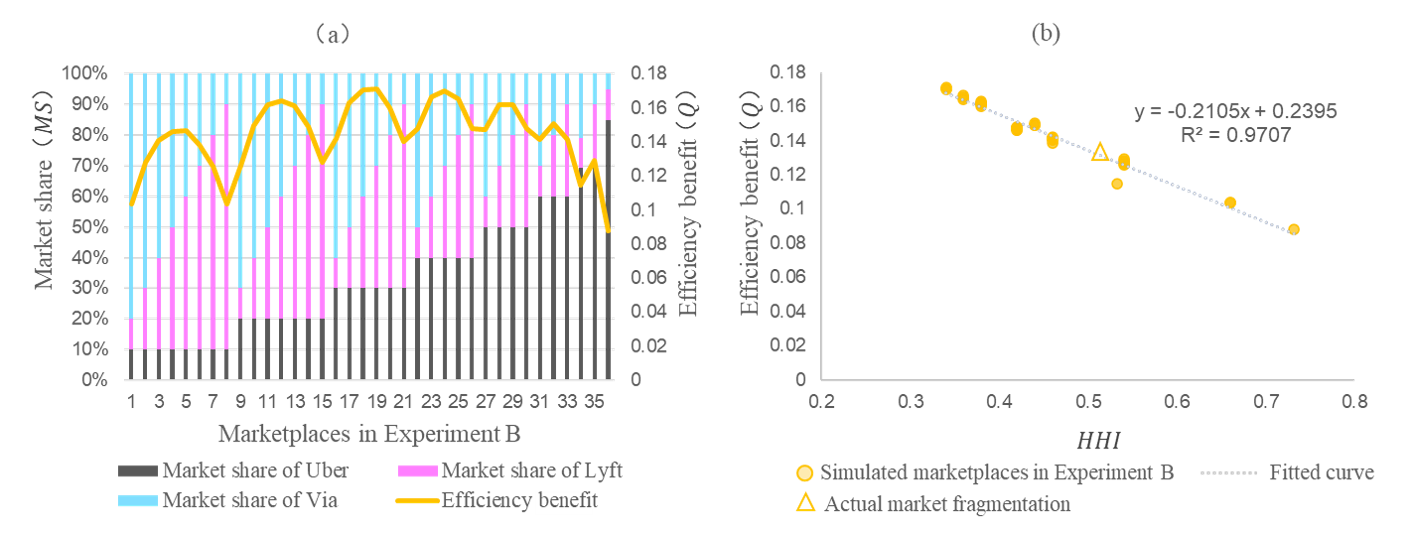}
  \caption{The impact of inter-TNC competition on the efficiency benefits of market integration}
  \label{fig:HHI}
\end{figure}

To investigate the effect of market competition, Experiment B maintains the existing demand concentration but changes the market shares of the 3 TNCs, simulating alternative markets with different competition intensities. Since the trips are randomly assigned, the simulated markets present negligible variation in modularity, ranging from 0.0005 to 0.0086. Therefore, the inter-TNC unevenness is assumed to be constant. In Fig.~\ref{fig:HHI}(a), the stacked bar chart represents various permutations of market splits among the 3 TNCs, and the curve represents the variation in potential efficiency benefits. It is found that a higher efficiency benefit can be achieved from market integration when market shares are more evenly distributed. Using $HHI$ to measure the intensity of market competition, the efficiency benefit of market integration is found to be in a negative linear relationship with the competition intensity (see Fig.~\ref{fig:HHI}(b)). The demand would be served less efficiently when the market is in intense competition, and as a result market integration would be more beneficial in such conditions.

It is noteworthy that the efficiency benefit from market integration varies greatly across TNCs. With the actual market shares in Manhattan (68.1\%:18.3\%:13.6\%), Uber, the largest TNC, would gain the least through market integration at 5.5\%. The individual benefit relates to the extent to which collaboration expands the shareability network for each TNC by creating more ridesharing opportunities. Through collaboration, smaller TNCs would gain more advantages, but larger TNCs provide more opportunities for smaller ones with fewer additional opportunities in return.

\begin{figure}[!ht]
  \centering
  \includegraphics[width=0.95\textwidth]{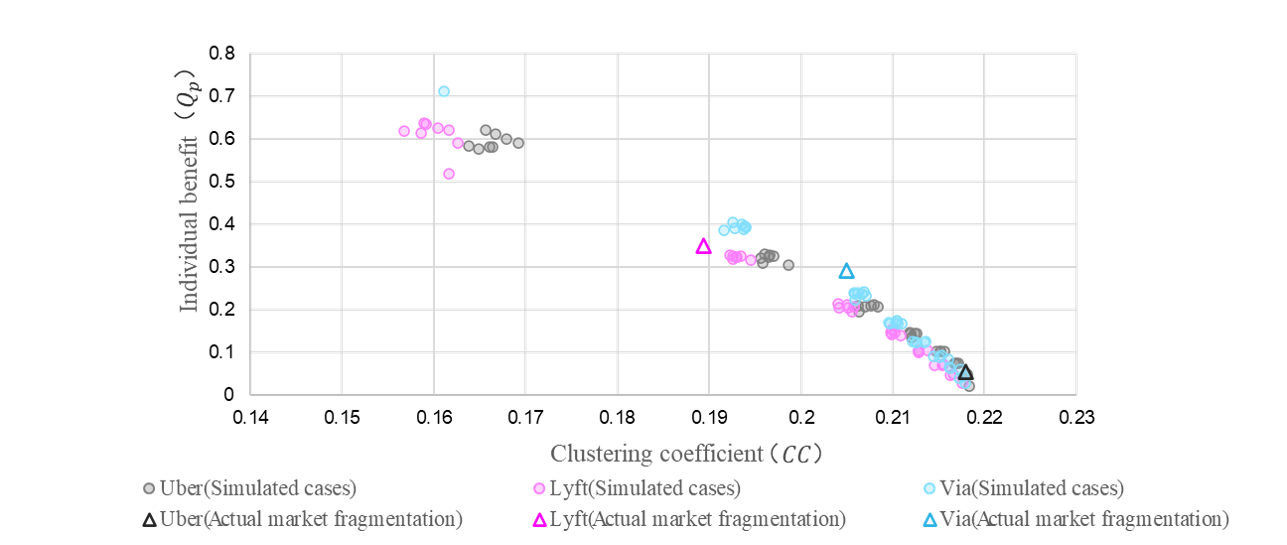}
  \caption{The impact of the intra-TNC unevenness on the individual benefits in market integration}
  \label{fig:individual_gain}
\end{figure}

Interestingly, Lyft has a larger market share than Via but benefits most from market integration. It implies that the market share of a TNC $p$ does not solely determines its individual benefit $Q_p$. It is likely that the specific spatiotemporal concentration of a TNC's demand pool also matters. For individual TNCs, an even and sparse trip distribution across space and time can limit the ridesharing efficiency. In such cases, market integration would be more beneficial, as it can expand the TNC's shareability network by providing more sharing opportunities. Therefore, we use the clustering coefficient ($CC$) to measure the extent of intra-TNC spatiotemporal concentration of trips. It turns out that, under the existing market split, Via has a lower market share, but a larger $CC$, than Lyft (see Table~\ref{table:characteristics}). In Experiment B, we calculate a $CC$ for each TNC based on each simulated market splits in Fig.~\ref{fig:HHI}(a). The results in Fig.~\ref{fig:individual_gain} show that the TNC with more clustered trip demand would generally have less to gain from market integration. As a result, the TNC with a larger clustering coefficient would have less incentive to collaborate with other TNCs. The uneven efficiency benefits to different TNCs can pose a serious barrier to market integration through collaboration between TNCs.

\subsection{Profit sharing analysis}

To address the issue with uneven individual benefits in VHT savings, a common strategy is to design a profit sharing scheme to maintain fairness in collaboration. The additional profits generated by cooperation between TNCs should be distributed according to each TNC's individual contribution. In this study, we propose to use the Shapley value to quantify the expected marginal contribution of each TNC in the integrated market, based on which we can allocate the profit to different TNCs. As shown in Table~\ref{table:Shapley}, $S$ are all possible alliances formed by any set of TNCs, including the non-TNC alliance \{$\varnothing$\}, single-TNC alliances for the fragmented market (i.e.\{Uber\}), permutations of bilateral collaborative alliances (i.e.\{Uber, Lyft\}), and the fully integrated alliance (i.e.\{Uber, Lyft, Via\}). The value of an alliance $val(S)$ is calculated by the potential VHT saving through a specific alliance $S$. The values of different alliances and the marginal contribution of each TNC are summarized in Table~\ref{table:Shapley}.

\begin{table}[ht!]
  \centering \footnotesize
  \caption{Calculation of the Shapley values}
    \begin{tabular}{c c c c c}
    \addlinespace
    \hline
    \ Alliance & Value of alliance &\multicolumn{3}{c}{Marginal contributions}\\
    \cmidrule(lr){3-5}
     ($S$) & ($val(S)$) & Uber & Lyft & Via\\
    \hline
    \{$\varnothing$\} & 0 & -& -& -\\
   \{Uber\} & 13865.1 & 13865.1 & -&- \\
   \{Lyft\} & 2909.0 &- & 2909.0 &- \\
   \{Via\} & 2865.1 & -&- & 2865.1 \\
   \{Uber, Lyft\} & 18150.0 & 14553.1 & 3596.9 &- \\
   \{Uber, Via\} & 17897.5 & 14448.8 & - & 3448.7 \\
   \{Lyft, Via\} & 6593.4 & - & 3318.7 & 3274.7\\
    \{Uber, Lyft, Via\} & 22246.4 & 14884.9 & 3754.8 & 3606.6 \\
    \hline
    Shapley value ($\phi$) & & 14884.9 & 3754.8 & 3606.6 \\
    \hline
    Profit shares & &  66.9\% & 16.9\% & 16.2\% \\
    \hline
    \end{tabular}
\label{table:Shapley}%
\end{table}%

Based on Eq.~\eqref{eq:Shapley}, we calculate the Shapley values for Uber, Lyft, and Via as $\phi=(14884.9,3754.8,3606.6)$, respectively. To ensure fairness, the alliance can allocate the additional profit as the result of market integration to Uber, Lyft, and Via in proportion to their Shapley values: (66.9\%:16.9\%:16.2\%). These proportions are close to but not exactly in line with the market shares in Table~\ref{table:characteristics}. 
Accordingly, Uber and Lyft should receive slightly lower shares of profit than their respective market shares, while Via's share of profit is higher than its market share. This result implies that the TNCs' individual contributions to the collaboration are not entirely determined by their market shares. Although Via serves the fewest trips among these three TNCs, it generates the largest average VHT savings per trip in both fragmented and integrated markets. The trips from Via are likely to be more profitable in ridesharing than other trips due to their characteristics or distribution, and therefore contribute the most to the integrated market at the per-trip level. 

As is shown in Table~\ref{table:Shapley}, the marginal contributions of TNCs in the integrated market are all larger than that in the fragmented or bilateral collaborative markets. It indicates that all TNCs can gain from collaborating with each other. Therefore, the three TNCs can form a stable and robust collaboration. Also, the result of the Shapley value calculation is unique so that the bargaining process among the TNCs can be avoided \citep{krajewska_horizontal_2008}.

\section{Discussion and Conclusion} 
In this study, we present a generalizable methodology to quantify the potential efficiency benefits of ridesharing market integration, as well as empirical findings based on a real-world fragmented market in Manhattan, NYC. Specifically, we construct shareability networks with the actual TNC trips from Uber, Lyft, and Via, and find that market integration would improve the ridesharing efficiency by 13.3\%, or roughly 5\% of the total VHT. It indicates that collaboration among the TNCs has substantial potential to dissolve the inefficiency induced by market fragmentation and mitigate the economic and environmental costs of traffic congestion. Therefore, the relevant authorities are alerted to take actions to facilitate market integration.

The use of real-world TNC trip data enables us to explore the impact of spatiotemporal unevenness in demand concentration across TNCs on ridesharing efficiency, an issue that has not been well studied in previous studies. The measurement of spatiotemporal unevenness can be challenging, especially when multiple TNCs are present in the market. In this work, we propose to use the modularity of the integrated shareability network to measure inter-TNC unevenness in the spatiotemporal demand distribution, and the clustering coefficient of the TNC-specific shareability network to measure the level of intra-TNC demand concentration. 

To understand how the efficiency benefit varies based on the market conditions, two experiments are conducted. 
Through Experiment A (\textit{Flexible Demand Patterns}), we fix the market split but vary demand concentration patterns. The results reveal how the efficiency benefits of market integration are related to both the demand density and market division (inter-TNC spatiotemporal unevenness). High demand density or significant market division would undermine the efficiency benefits of market integration. Significant market division occurs when competitors negotiate and divide markets by geographic areas or target customers. In the real world, moderate market division is common, as some smaller/newer TNCs may attempt to mitigate direct competition with established TNCs and instead target a niche market. Eq.~\eqref{eq:market} quantifies the extent to which demand density and market division affect the efficiency benefits, which could assist policymakers to assess the need for facilitating market integration based on real demand distribution.

To further explore how the efficiency benefits of market integration vary with competitive intensities, Experiment B (\textit{Flexible Market Splits}) simulates markets with different market shares across TNCs. It is found that a larger efficiency benefit could be achieved from integration when the market shares are more evenly distributed. While previous studies on the impact of market shares are limited to duopoly markets \citep{zhang_economies_2022,guo_dissolving_2022}, we use $HHI$ to describe the extent of market fragmentation and competition, enabling the approach to be applied to multi-TNC markets. Combining the two experimental analyses, policymakers can adapt the proposed framework to evaluate the efficiency benefits of market integration in other cities with various market conditions.

Focusing on the perspective of individual TNCs, we find that their individual efficiency benefits vary greatly and can be determined by factors beyond just the market share. For example, Lyft holds a larger market share than Via in Manhattan, but also has more to gain from market integration. It is revealed that the intra-TNC unevenness also plays a more important role in determining individual benefits. Simulation results show that a TNC's individual benefit is negatively related to its clustering coefficient. The uneven benefits across TNCs suggest a barrier to collaboration. To ensure fairness, we also propose to utilize the Shapley value to evaluate the marginal contributions by each TNC and design profit-sharing schemes accordingly.

In practice, collaboration among TNCs can represent a substantial threat of monopoly, as the alliance could easily manipulate the price. To deal with it, the government could assume the role of a central broker to supervise collaboration by enforcing a price ceiling or a maximum rate of return. It is in the interests of both drivers and passengers to prevent TNCs from abusing market power for higher profits. From the perspective of TNCs, the public sector involvement also contributes to ensuring fairness and transparency, which will help build trust among TNCs.
Incentives can be provided to enhance their willingness to collaborate. For example, the government may introduce congestion charging reduction or taxes reliefs for TNCs involved in the alliance. This study provides a general approach for the market regulators and business alliances to evaluate and monitor the fragmented market efficiency and dynamically adjust their strategies, incentives, and profit-sharing schemes based on updated trip records for greater market efficiency while ensuring fairness.

There are several limitations in this study. First, due to the lack of exact locations in TNC trip records, the spatial concentration of demand may be inaccurate, leading to bias in quantifying spatiotemporal unevenness and efficiency benefits. Also, we assume an average speed to the street network, which neglects congestion effects in route selection and ridesharing matching. These issues all stem from data limitations, and should be addressable when more detailed data become available. Second, as for the method, the shareability network is built following the Oracle model in \cite{santi_quantifying_2014}. This method assumes that the travel demand throughout the day is already known and thus provides the optimal matching in terms of overall travel cost savings. However, actual operations of ridesharing services often require real-time computation and matching with imperfect knowledge about the future \citep{schreieck_matching_2016,tafreshian_trip-based_2020}. As a result, the estimated ridesharing efficiency could be slightly higher than real-world systems. Therefore, the proposed approach is more suitable for post hoc market analysis or policy evaluation, instead of real-time operations management. Third, compared to the previous studies based on the equilibrium models \citep{ZHOU2022103851, zhou_competitive_2022b}, the simulation experiments in this study cannot account for the TNCs' behaviors in response to the different market fragmentation levels. For instance, the TNCs may adjust their operating strategies in pricing, spatiotemporal concentration, and ridesharing thresholds, which may, in return, affects the market efficiency and individual benefits. In addition to the impact of spatiotemporal concentration of demand, future studies on the factors influencing fragmented market efficiency could also consider the TNCs' various business strategies in terms of pricing, driver wages, etc.

\section*{Acknowledgements}
This research is partly supported by the Seed Funding for Strategic Interdisciplinary Research Scheme (102010057) at the University of Hong Kong.

\appendix

\bibliographystyle{model5-names2}\biboptions{authoryear}
\bibliography{main}







\end{document}